\documentclass[pre,twocolumn,showpacs]{revtex4}
\usepackage{epsfig}

\begin{document}

\preprint{MZ-TH/08-11}

\title{Short chaotic strings and their behaviour in the scaling region}

\author{Stefan Groote}
\affiliation{Teoreetilise F\"u\"usika Instituut, Tartu \"Ulikool,
T\"ahe 4, 51010 Tartu, Estonia\\
and Institut f\"ur Physik der Universit\"at Mainz,
Staudingerweg 7, 55099 Mainz, Germany}
\email{groote@thep.physik.uni-mainz.de}
\author{Hardi Veerm\"ae}
\affiliation{Teoreetilise F\"u\"usika Instituut, Tartu \"Ulikool,
T\"ahe 4, 51010 Tartu, Estonia}
\email{hardi27@ut.ee}

\vspace{2cm}

\begin{abstract}
Coupled map lattices are a paradigm of higher-dimensional dynamical systems
exhibiting spatio-temporal chaos. A special case of non-hyperbolic maps are
one-dimensional map lattices of coupled Chebyshev maps with periodic boundary
conditions, called chaotic strings. In this short note we show that the fine
structure of the self energy of this chaotic string in the scaling region
(i.e.\ for very small coupling) is retained if we reduce the length of the
string to three lattice points.
\end{abstract}


\maketitle

\section{Introduction}
Coupled map lattices (CMLs) as introduced by Kaneko and Kapral~\cite{kaneko,%
kapral} are a paradigm of higher-dimensional dynamical systems exhibiting
spatio-temporal chaotic behaviour. There is a variety of applications for CMLs
to model hydrodynamical flows, turbulence, chemical reactions, biological
systems, and quantum field theories (see e.g.\ reviews in~\cite{kanekobook,%
beckbook}). For hyperbolic maps with slope of the local maps always larger
than one a variety of analytical results exists~\cite{baladi,jarvenpaa,keller,%
bunimovich}. However, promising results have been achieved also for the
physically more interesting non-hyperbolic maps~\cite{chate,ding,ruffo,mackey,%
beck,gallas,grootes,grootel}.

Local non-hyperbolic maps given by $N$-th order Che\-byshev polynomials promise
to answer a long-standing problem of elementary particle physics, namely the
mass spectrum of elementary particles. With surprizing evidence it could be
shown that the self-energy of the chaotic string, i.e.\ a one-dimensional
lattice based on diffusely coupled Chebyshev maps, shows a fine structure with
minima that can be related to the fundamental mass parameters of the standard
model of elementary particle physics~\cite{beckbook,beck}. It was recently
shown that this fine structure disappears if we use a lattice of more than one
dimension~\cite{maher}. In addition the chaotic string can serve as a useful
model for vacuum fluctuations and dark energy in the universe~\cite{dark}.

It is interesting to know whether under special circumstances a long
one-dimensional lattice with periodic boundary conditions can be replaced by a
short one. Dettmann {\it et al.\/} studied a string of length $L=2$, i.e.\ the
case of two coupled Chebyshev maps~\cite{dettmann} using periodic orbit theory.
In this note we examine short strings with periodic boundary conditions
(closed strings) as well as strings with open ends (open strings) and fixed
ends (fixed strings). Our central result is that closed strings of length
$L\ge 3$ as well as other short string configurations lead to the same self
energy fine structure in the scaling region (i.e.\ for the coupling close to
zero) as for the case of maps with rather long lengths ($L=10\,000$). This
fact supports hopes for understanding the reason for the fine structure in the
scaling region.

\section{Approaching the scaling region}
In this note we consider the CML given by the coupled Chebyshev map system
\begin{equation}\label{CML}
\phi_i^{n+1}=(1-a)T_2(\phi_i^n)+\frac a2\left(T_2(\phi_{i+1}^n)
  +T_2(\phi_{i-1}^n)\right)
\end{equation}
which is denoted as 2A-string in Ref.~\cite{beckbook}. The index $n$ counts
the iterations while $i$ names the position on the one-dimensional string. We
are calling these numbers the temporal and the spatial position. Periodic
boundary conditions close the string by implying $\phi_{L+1}^n=\phi_1^n$ where
$L$ is the string length. $a$ is the coupling and $T_2(\phi)$ is the second
order Chebyshev polynomial
\begin{equation}
T_2(\phi)=2\phi^2-1.
\end{equation}
It can be shown that in the scaling region ($a$ small) the 2B-string (with the
last two Chebyshev polynomials in Eq.~(\ref{CML}) replaced by identities) has
the same behaviour.

\begin{figure}[ht]
\epsfig{figure=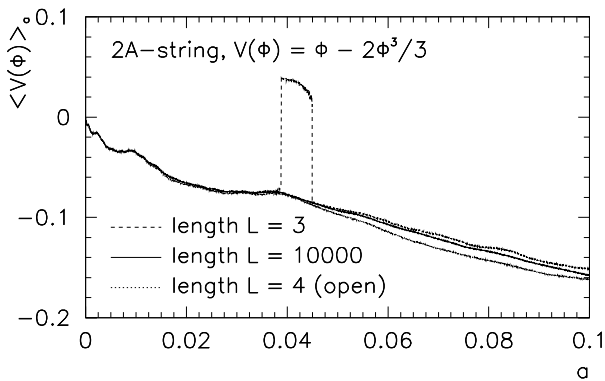, scale=0.5}
\caption{\label{self2a34}Self energy of a closed 2A-string of length
$L=10\,000$ (solid line), a closed 2A-string of length $L=3$ (dashed), and an
open 2A-string of length $L=4$ (dotted) for $a\in[0,1]$}
\end{figure}

The self energy of the chaotic string is the temporal and spatial average
\begin{equation}
\langle V(\phi)\rangle_a=\langle\phi-\frac23\phi^3\rangle.
\end{equation}
It is obtained by performing the map~(\ref{CML}) iteratively and taking the
average over all values $\phi_i^n$ on the string. Obviously, the precision of
the result is growing with the product of the number of iterations and the
length of the (closed) string. The self energy depends on the coupling $a$ in
a non-obvious way. In Fig.~\ref{self2a34} we show the self energy for a closed
string of length $L=10\,000$, a closed string of length $L=3$, and an open
string of length $L=4$.
The latter is given by the map~(\ref{CML}) for $i=2,3$ combined with
$\phi_1^{n+1}=T_2(\phi_1^n)$ and $\phi_4^{n+1}=T_2(\phi_4^n)$. The boundary
values are not influenced but themselves influence the central value. Because
of the ergodicity of the Chebyshev map they are distributed according to the
invariant probability density\footnote{Note that a possible alternative to
this prescription for the open string is to replace the deterministic
evoluation given by the Chebyshev map by a stochastic variable without time
correlation, distributed according to $\rho_0(\phi)$.} 
\begin{equation}
\rho_0(\phi)=\frac1{\pi\sqrt{1-\phi^2}}.
\end{equation}
Looking at Fig.~\ref{self2a34} we see that close to $a=0$ the differences
between the three results seem to vanish. For larger values of $a$ the self
energy of the closed string of length $L=3$ disagrees to positive values, the
self energy of the open string to negative values. The noticeable disagreement
between the closed strings for length $L=10\,000$ and $L=3$ close to $a=0.04$
is due to the formation of stable orbits on the short string. A detailed
analysis of the interval $a\in[0.039,0.043]$ shows that stable orbits of
different periods interchange with chaotic behaviour. However, we do not dwell
on this point here because the value $a=0.04$ is still considered to be
outside of the scaling region.

\section{Fine structure of the self energy}
\begin{figure}[ht]
\epsfig{figure=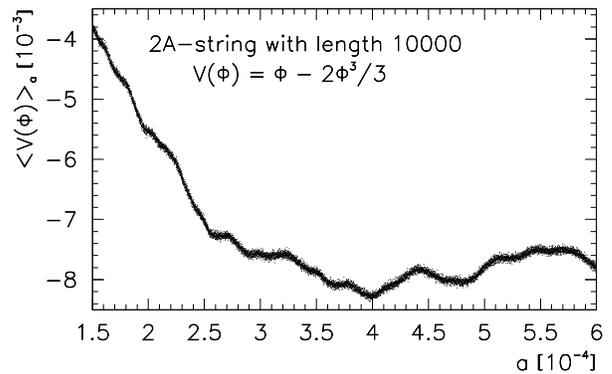, scale=0.5}
\caption{\label{self2a}Self energy of the 2A-string of length $L=10\,000$ in
the range $a\in[0.00015,0.0006]$ after $10\,000$ iterations}
\end{figure}

The name ``scaling region'' for the region close to $a=0$ is due to the fact
that we observe the scaling behaviour
\begin{equation}
\langle V(\phi)\rangle_a-\langle V(\phi)\rangle_0=f^{(N)}(\ln a)\sqrt a
\end{equation}
where $f^{(N)}$ is a periodic function of $\ln a$ with period
$\ln N^2$~\cite{beckbook}. Using perturbative methods, this scaling behaviour
could be shown in Refs.~\cite{grootes,grootel}. Looking at a single period for
the 2A-string (i.e.\ $N=2$) we observe a fine structure of the self energy
which is not yet fully understood. However, the local minima of the self
energy could be shown to be related to Yukawa and gravitational couplings for
all quark and lepton flavours modulo $N^2=4$. Therefore, the fine structure of
the self energy as shown in Fig.~\ref{self2a} is of central interest for us.

\section{Closed string of length $L=2$}
\begin{figure}[t]
\epsfig{figure=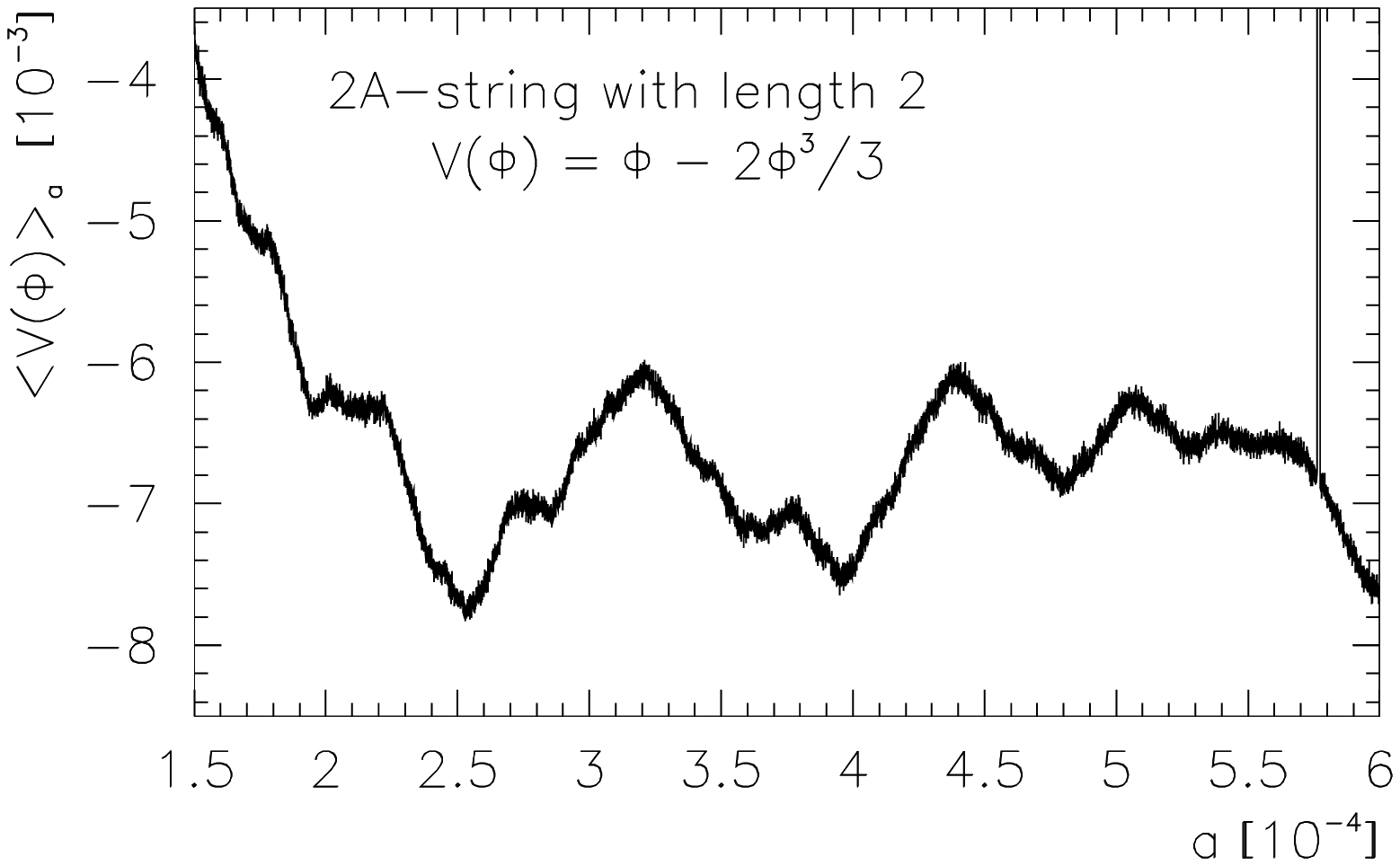, scale=0.5}
\caption{\label{self2a2}Self energy of the 2A-string of length $L=2$ with
periodic boundary conditions in the range $a\in[0.00015,0.0006]$, the number
of iterations being $3\times 10^7$}
\vspace{12pt}
\epsfig{figure=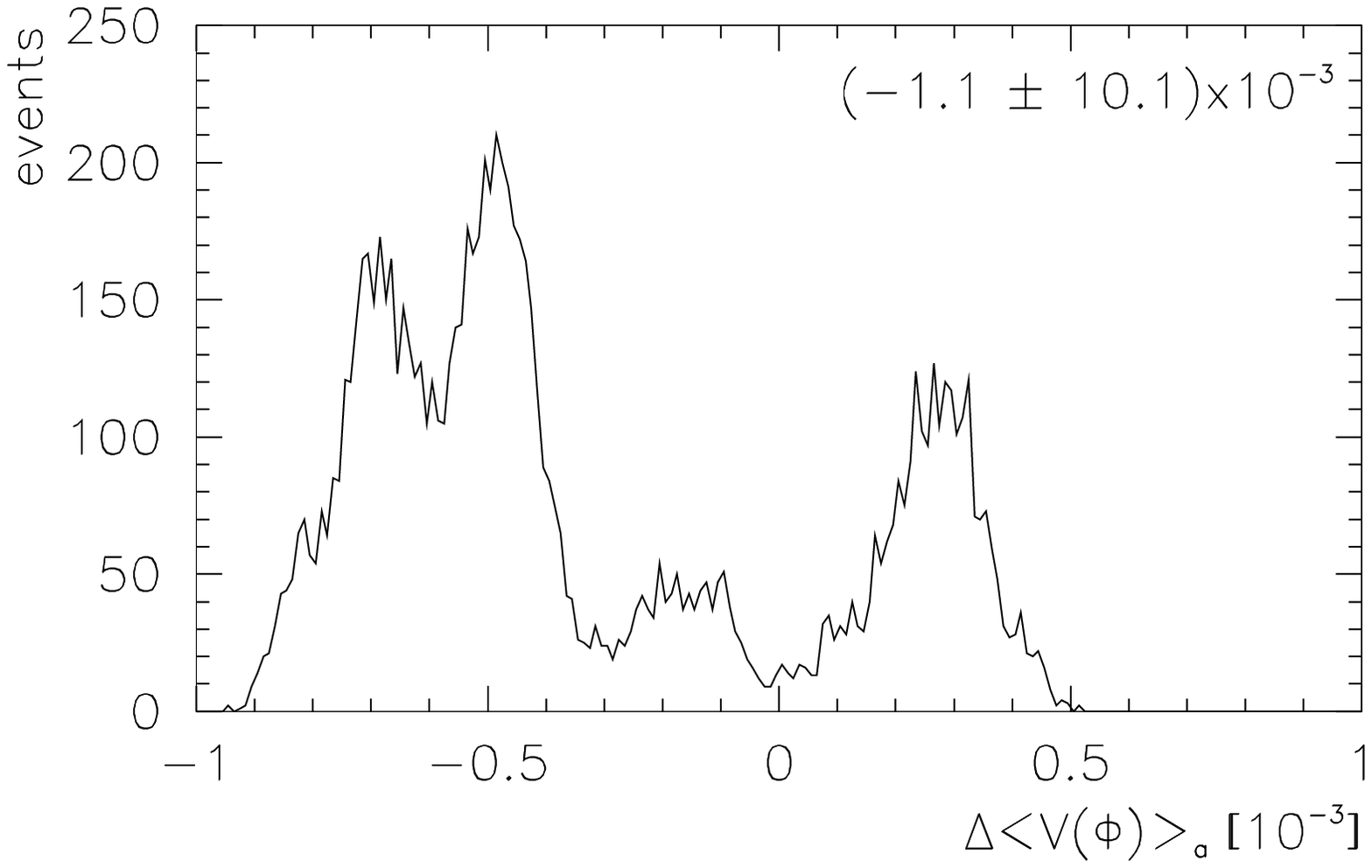, scale=0.5}
\caption{\label{self2a2d}Distribution function for the difference values of
the self energies of the closed strings for $L=2$ and $L=10\,000$ mediated
over the intervall $a\in[0.00015,0.0006]$}
\end{figure}

If we reduce the string length to two points as in Ref.~\cite{dettmann} by
keeping the periodic boundary conditions we observe a quite different shape of
the self energy (Fig.~\ref{self2a2}). A fine structure is still visible but is
different from the one for the long string. Relative minima are reallocated
and, therefore, are not related to physical parameter values. The self-similar
structure is much more evident as in the other cases. The enhancement close to
$a\approx 0.000577$ (similar to the one for the closed string of length $L=3$
near $a=0.04$) is caused by stable orbits interchanged with chaotic behaviour.
At $a=0.0005768$ for instance we find an orbit of period $6$ which reads
\begin{eqnarray}
+0.7070&&+0.9954\nonumber\\
+0.0003&&+0.9811\nonumber\\
-0.9989&&+0.9238\nonumber\\
+0.9954&&+0.7070\nonumber\\
+0.9811&&+0.0003\nonumber\\
+0.9238&&-0.9989
\end{eqnarray}
The deviation between the fine structures in Figs.~\ref{self2a}
and~\ref{self2a2} can be measured by the mean value of the differences of the
self energy value mediated over $a\in[0.00015,0.00060]$ which is given by
$(-1.2\pm 10.1)\times 10^{-3}$. The distribution function of this difference
measured on the interval $a\in[0.00015,0.00060]$ differs very much from a
Gaussian distribution function (cf.\ Fig.~\ref{self2a2d}). While the large
standard deviation is due to the orbit region, the shape of the distribution
function and the large value of the order of $10^{-3}$ for the mean value
indicates the visible deviation of the two fine structures.

\section{Closed string of length $L=3$}
\begin{figure}[t]
\epsfig{figure=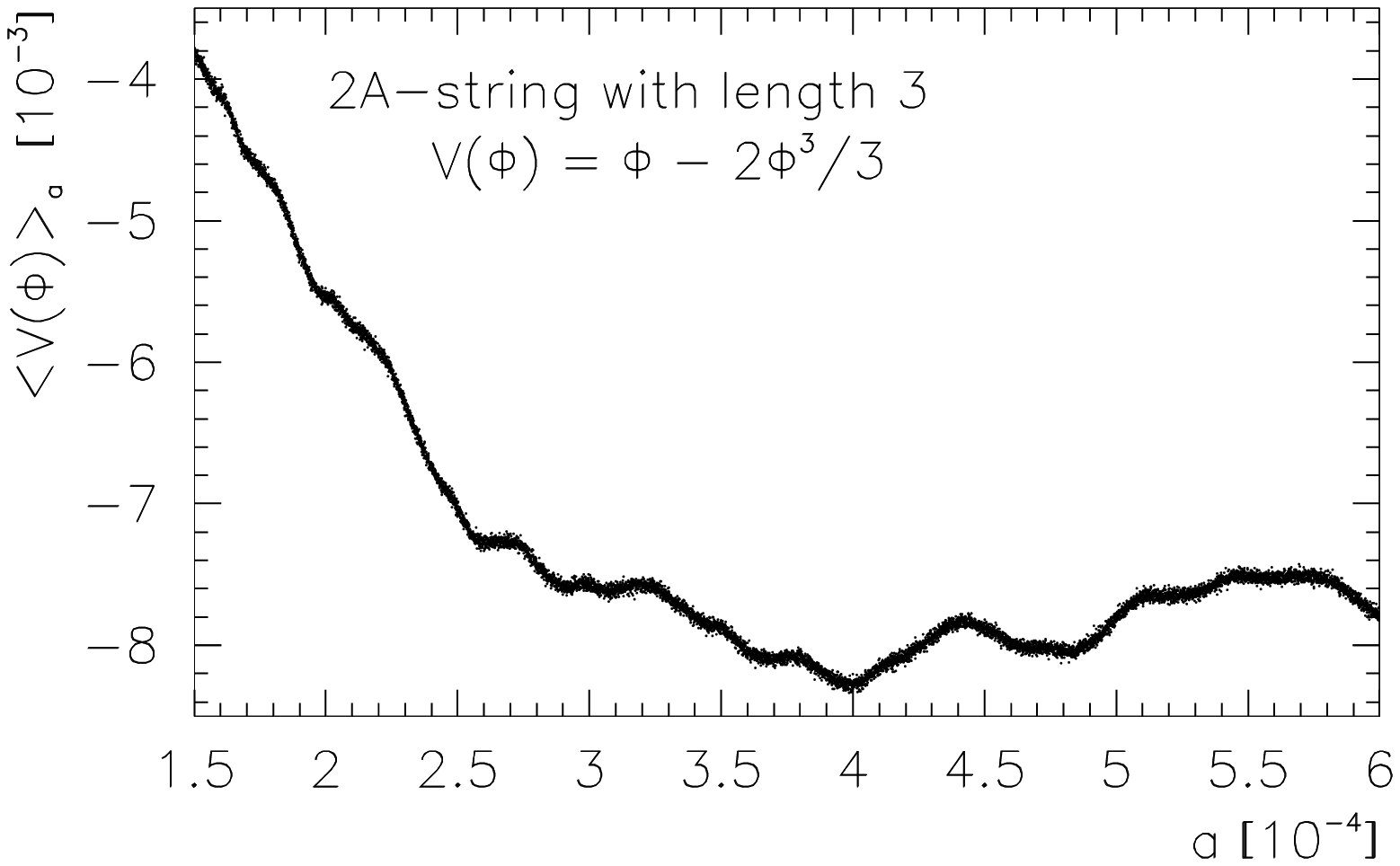, scale=0.5}
\caption{\label{self2a3}Self energy of the 2A-string of length $L=3$ with
periodic boundary conditions in the range $a\in[0.00015,0.0006]$, the number
of iterations being $6\times 10^7$}
\vspace{12pt}
\epsfig{figure=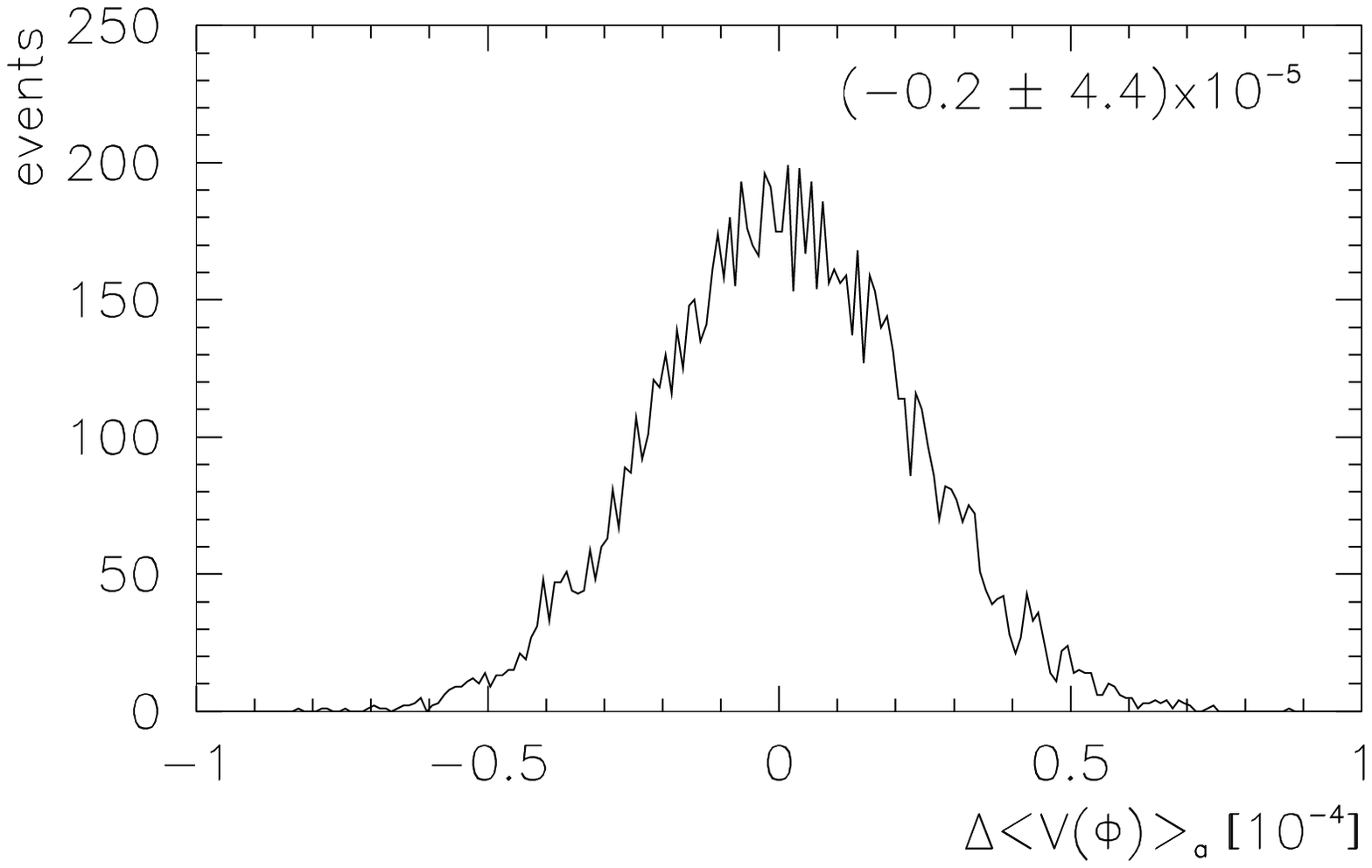, scale=0.5}
\caption{\label{self2a3d}Distribution function for the difference values of
the self energies of the closed strings for $L=3$ and $L=10\,000$ mediated
over the intervall $a\in[0.00015,0.0006]$}
\end{figure}

If we are looking at the closed string of length $L=3$, we observe that the
fine structure is recovered (Fig.~\ref{self2a3}). The mean value of the
differences $(0.2\pm 4.4)\times 10^{-5}$ is quite small. The standard
deviation is due to statistical errors which could be reduced by calculating
the two fine structures to higher accuracy. Finally, the distribution function
is very much of the shape of a Gaussian curve (Fig.~\ref{self2a3d}). The same
is true for other closed strings of short length, for instance for a string
consisting of $L=4$ or $L=6$ points (mean difference values are
$(0.3\pm 6.6)\times 10^{-5}$ and $(0.2\pm 5.7)\times 10^{-5}$, resp.).

\begin{figure}[t]
\epsfig{figure=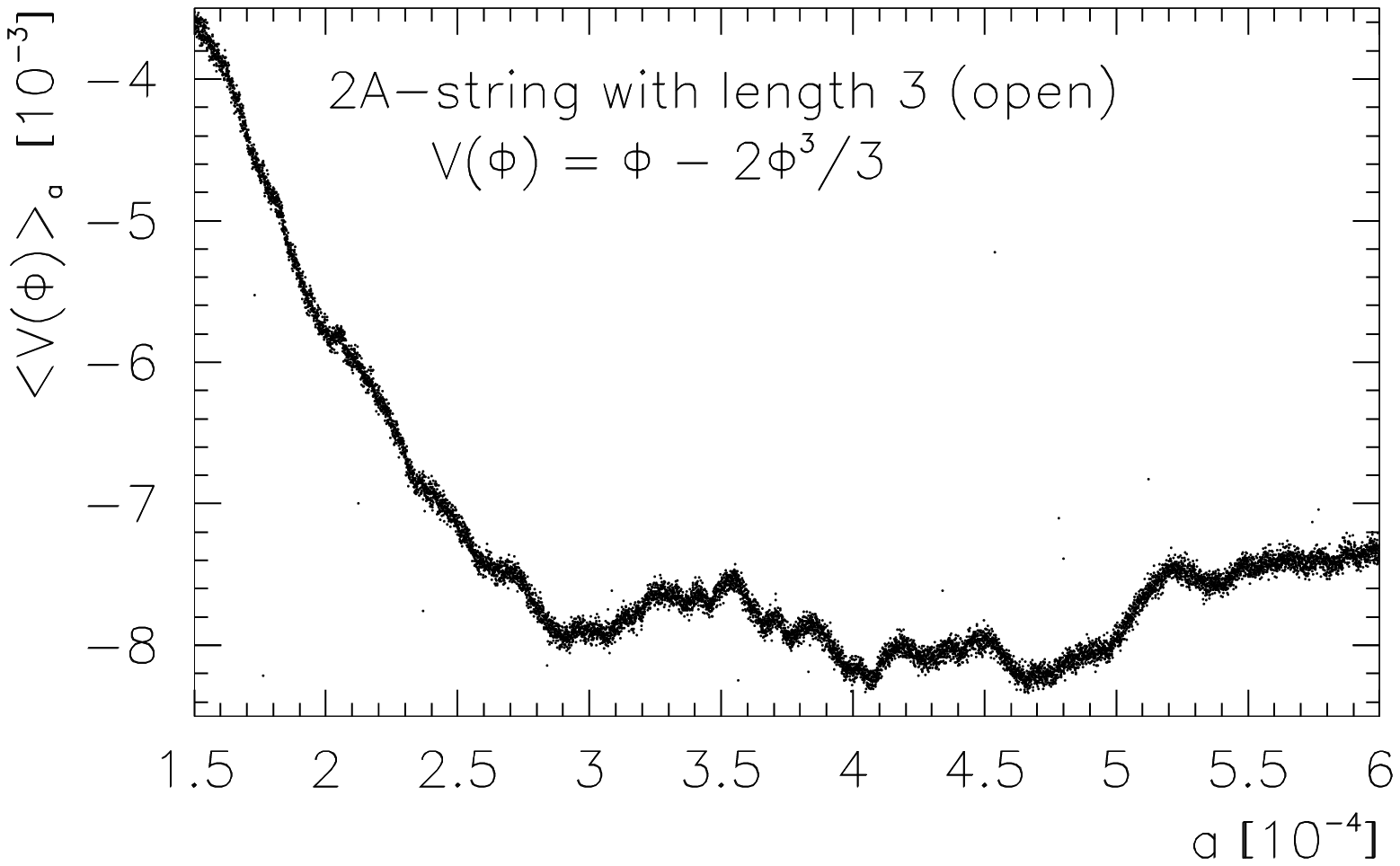, scale=0.5}
\caption{\label{self2a3o}Self energy of the 2A-string of length $L=3$ with
open ends in the range $a\in[0.00015,0.0006]$ after $6\times 10^7$ iterations}
\vspace{12pt}
\epsfig{figure=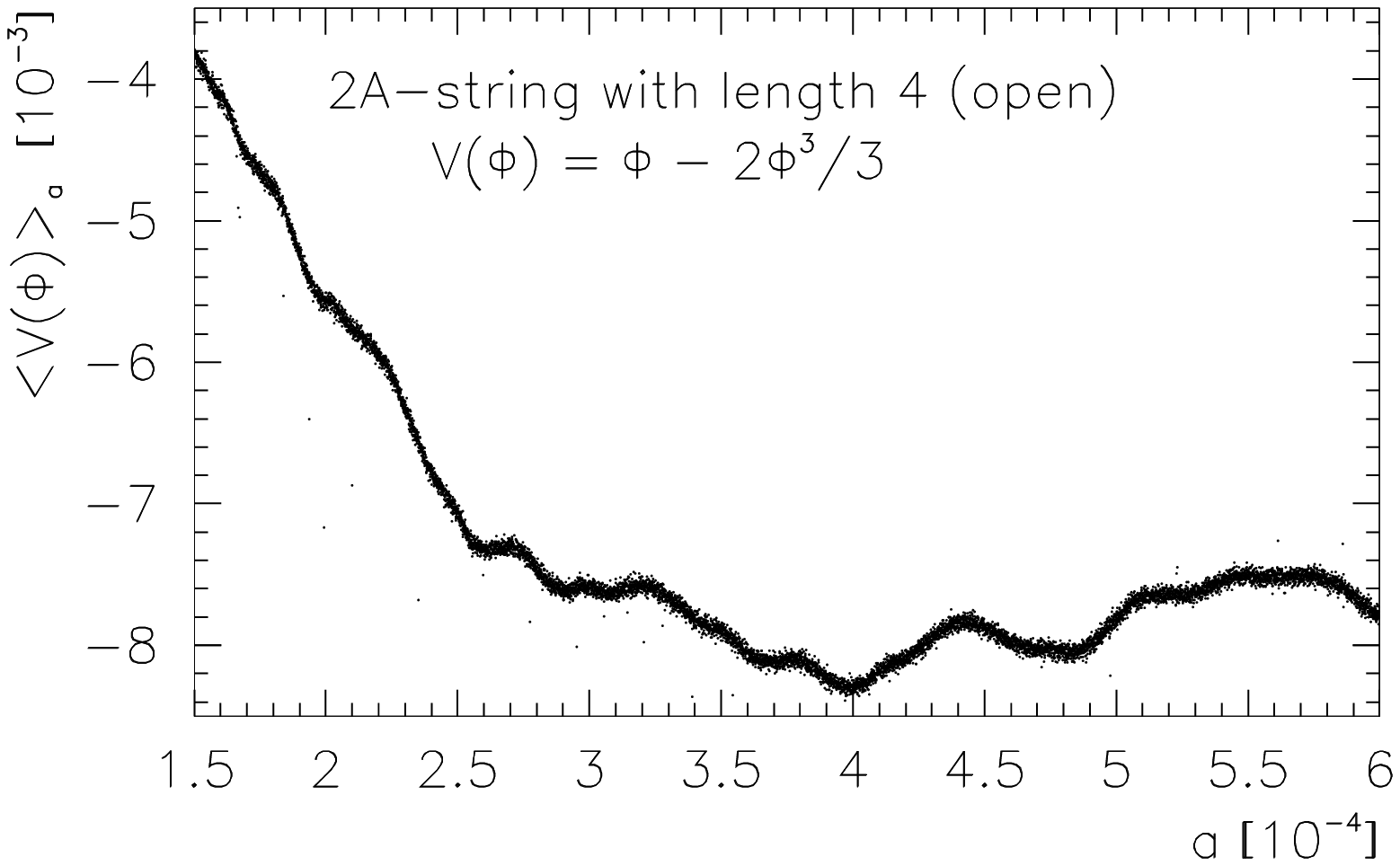, scale=0.5}
\caption{\label{self2a4o}Self energy of the 2A-string of length $L=4$ with
open ends in the range $a\in[0.00015,0.0006]$ after $6\times 10^7$ iterations}
\end{figure}

\section{Short open strings}
If we replace the perodic boundary conditions by the open string boundary
conditions (map~(\ref{CML}) for $i\ne 1,L$ together with
$\phi_1^{n+1}=T_2(\phi_1^n)$ and $\phi_L^{n+1}=T_2(\phi_L^n)$), we obtain a
different picture (cf.\ Fig.~\ref{self2a3o}). The open string of length $L=3$
corresponds to the perturbative settings given in Refs.~\cite{grootes,grootel}.
A fine structure is visible but is again different from the one for the closed
string. This fact is indicated also by the mean difference
$(1.6\pm 1.8)\times 10^{-3}$ and the distribution function which is divided up
into two maxima. However, if we use an open string of length $L=4$
(Fig.~\ref{self2a4o}), the original pattern of Fig.~\ref{self2a} is recovered
again (mean difference $(1.6\pm 6.2)\times 10^{-5}$). In both cases we observe
some scattered points which deviate from the main distribution. An analysis
shows that unlike the case of closed strings, these points do not represent
stable orbits. Instead, their occurence depends on the starting conditions and
would vanish if we sum over a sufficient large set of random starting values.

\section{Short fixed strings}
\begin{figure}[b]
\epsfig{figure=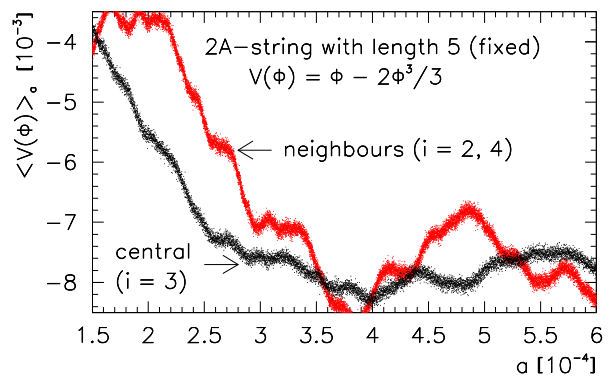, scale=0.5}
\caption{\label{self2a5f}Self energy of the 2A-string of length $L=5$ with
fixed ends in the range $a\in[0.00015,0.0006]$ after $3\times 10^7$ iterations
for the central point $i=3$ and the neighboured points ($i=2,4$)}
\end{figure}

A final analysis is devoted to a one-dimensional map lattice with border
values set to $\phi_1^n=\phi_L^n=0$ (fixed string). The first reoccurence of
the characteristic fine structure is obtained for the fixed string of length
$L=5$ for the central value of this string ($i=3$). The neighboured points
($i=2,4$) also show a fine structure but more of the kind we were used to
from the closed string of length $L=2$. Both fine structures are shown in
Fig.~\ref{self2a5f}.

\section{Conclusions}
We have compared the self energy of the 2A-string of length $L=10\,000$
and periodic boundary conditions (closed string) with the same closed string
with short lattice sizes. It turns out that for $L\ge 3$ the fine structure of
the self energy in the scaling region close to $a=0$ can be reproduced.
Using an open string of short length it figures out that the fine structure is
reproduced for a string of length $L\ge 4$. Finally, the central points of a
string of length $L\ge 5$ with fixed ends shows again the same fine structure
of the self energy. We conclude that the closed string of length $L=3$, the
open string of length $L=4$, or the fixed string of length $L=5$ are minimal
requirements for understanding the origin of the fine structure of the self
energy dependence on the coupling $a$ in the scaling region.

\begin{acknowledgments}
We thank C.~Beck for discussion and encouragement. This work is supported in
part by the Estonian target financed project No.~0182647s04 and by the
Estonian Science Foundation under grant No.~6216. S.G.\ also acknowledges
support from a grant of the Deutsche Forschungsgemeinschaft (DFG) for staying
at Mainz University as guest scientist for a couple of months.
\end{acknowledgments}

\advance\textheight by12pt

\end{document}